\documentclass[10pt,conference]{IEEEtran}

\IEEEoverridecommandlockouts
\newcommand{\channel}{{OTDR channel }}
\newcommand{\fN}{\overrightarrow{\mathcal W}}
\newcommand{\bN}{\overleftarrow{\mathcal W}}
\newcommand{\err}{\mathrm{err}}

\usepackage{cite}
\usepackage{amsmath,amssymb,amsfonts}
\usepackage{algorithmic}
\usepackage{graphicx}
\usepackage{textcomp}
\usepackage{xcolor}
\usepackage{graphicx,ragged2e}
\usepackage{physics}
\usepackage{float}
\usepackage{subfiles}
\usepackage{bbm}
\usepackage{adjustbox} 
\usepackage{dsfont}
\usepackage{xcolor} 
\usepackage[rightcaption]{sidecap}
\usepackage{glossaries}
\usepackage{wrapfig}
\newtheorem{theorem}{Theorem}
\newtheorem{definition}{Definition}
\newtheorem{lemma}{Lemma}

\newacronym{idr}{IDR}{Intrusion Detection Rate}
\newacronym{dtr}{DTR}{Data Transmission Rate}
\newacronym{otdr}{OTDR}{Optical Time Domain Reflectometer}
\newacronym{lcst}{LCST}{Pure-Loss Coherent State Transmitter}
\def\BibTeX{{\rm B\kern-.05em{\sc i\kern-.025em b}\kern-.08em
    T\kern-.1667em\lower.7ex\hbox{E}\kern-.125emX}}

\begin{document}

\title{Joint Communication and Eavesdropper Detection on the Lossy Bosonic Channel
\thanks{This work was financed by the DFG via grants NO 1129/2-1 and NO 1129/4-1 and by the Federal Ministry of Education and Research of Germany in the program of ``Souver\"an. Digital. Vernetzt.''. Joint project 6G-life, project identification number: 16KISK002 and via grants 16KISQ077 and 16KISQ168 and by the European Union in the project Q-net-Q. Further the support of the Munich Quantum Valley (MQV) which is supported by the Bavarian state government with funds from the Hightech Agenda Bayern Plus and Munich Center for Quantum Science and Technology is gratefully acknowledged. JN acknowledges inspiring discussions with N. Hanik at the Workshop on Entanglement-ASsisted Communication Networks (EACN) 2023.}
}

\author{
\IEEEauthorblockN{Pere Munar-Vallespir, Janis Nötzel, Florian Seitz}
\IEEEauthorblockA{\textit{Emmy Noether Group Theoretical Quantum System Design} \\
\textit{Technische Universität München}\\
Munich \\
\{pere.munar,janis.noetzel,flo.seitz\}@tum.de}
}

\maketitle

\begin{abstract}
    We study the problem of joint communication and detection of wiretapping on an optical fiber from a quantum perspective. Our system model describes a communication link that is capable of transmitting data under normal operating conditions and raising a warning at the transmitter side in case of eavesdropping. It contributes to a unified modelling approach, based on which established quantum communication paradigms like quantum key distribution can be compared to other approaches offering similar functionality. 
\end{abstract}


\section{Introduction}
    In this paper, we study the role of quantum technology in communication beyond quantum key distribution (QKD). Since detecting an eavesdropper is typically part of a QKD protocol, we ask whether this specific aspect may also be embedded into the data transmission protocol. Our problem definition thus defines a system that reliably transmits data under normal operating conditions and raises a warning if eavesdropping is detected.
    
    Integrated communication and sensing platforms are important current research topics. From the network research perspective, current efforts can be traced back to studies made during investigations on 5G systems \cite{wymeersch2021JCAS}, which carried over to 6G systems \cite{wymeersch2024JCAS}, resulting in a recent demonstration of a fully integrated platform \cite{andrei2024digital}. We follow this trend but focus on optical fiber.  

    At the same time, quantum technology promises advanced sensing methods going strictly beyond the available methods using conventional technology \cite{degenQuantumSensing}. Specifically in the domain of communication, the aspect of detection and localization of a possible eavesdropper has been studied from an applied perspective in the recent work \cite{popp2023eavesdropper}, where the system concept is to identify the presence of an eavesdropper via a QKD protocol and then identify the location of the eavesdropper using frequency-modulated Brillouin optical correlation domain analysis. Another concept called ``Quantum Alarm'' has been investigated in \cite{gong2020}, where an optical fiber data transmission system multiplexes the classical data signal stream with specific signals designed for eavesdropper detection. In particular, the work \cite{gong2020} discussed the possibility of integrating their quantum alarm system into existing fiber networks. The first work on an information theoretical approach to quantum joint communication and sensing was done by Want et al.\cite{Wang_2022}, where classical data is sent on a memory-less classical-quantum channel in finite dimensions and, at the same time, quantum hypothesis testing over a finite set was performed on a backward channel. In the following, we present a similar work but on a channel with memory and in an infinite dimensional Hilbert space. 
    
    However, the question arises how to quantify the range of possible operating points of a system that transmits classical data while raising an alarm when an eavesdropping attempt is detected. With this work, we provide a starting point for such analysis. We start with a model definition that is inspired by the concept of the (quantum) compound channel \cite{bb07,hayashiUniversal}, which has recently also been investigated from the perspective of optical systems \cite{cnr}. In such a model, the legitimate parties cannot be sure which channel they are transmitting over. Therefore, they are forced to use codes that work simultaneously for a large class of possible channels. We deviate from this earlier work by defining a specific channel state $s=0$, which defines the normal operating conditions. All other states $s\neq0$ define deviations from the normal conditions. In general, there are three possibilities for the communication system to carry out this task: By letting the transmitter, the receiver, or both carry out the task of detection. The previous works \cite{popp2023eavesdropper,gong2020} used the third option. In our approach, we use a purely sender-side method. In applications, such a decision may reduce system complexity and latency. We motivate sender-side channel state information via back-scattering. Traditionally, this method is also used in Optical Time-Domain Reflectometry (OTDR) and has also been used to demonstrate intrusion detection systems \cite{juarezDistributedIntrusionSensor,koyamadaDistributedSensingOTDR}. The question of how to model both the communication and the sensing aspect is approached in our work by introducing tuples $(D, R)$ modeling the detection and data transmission performance. We characterize several such operating points for the particular model derived in this work. We point out the performance gains that can be harvested using quantum detection methods over comparable conventional implementations. In contrast to \cite{popp2023eavesdropper}, our model does not include a step for identifying the location of the eavesdropper.
\section{Notation and Definitions}
    \subsubsection{Notation}
    We denote as $\ket{n}\forall n\in\mathbb N$ the photon number- or Fock-basis. $\ket{x^n}\forall x^n\in\mathbb C^n$ refers to $\bigotimes_{i=1}^n \ket{x_i}$ where $\ket{x_i}$ are coherent states. A coherent state is defined as $\ket{x_i} = e^{-|x_i|^2/2}\sum_{n=0}^\infty \tfrac{(x_i)^n}{\sqrt{n!}}\ket{n}$. We denote by $D(\beta^n)$ the displacement operator, which is acts as $D(\beta^n)\ket{\alpha^n} = \ket{\alpha^n+\beta^n}$. The Holevo information of a channel $\mathcal W\in C(\mathcal X,\mathcal K)$ and and ensemble $\{p_x,\rho_x\}_{x\in\mathcal X}$ is written as $\chi(p;\mathcal W)$. In general, it describes the rate of a classical-quantum channel; thus, the capacity can be calculated as  $C(\mathcal{W}) = \max_{\rho_x, p_x} \chi(p;\mathcal W)$.
    $\log$ denotes the logarithm in base 2 and $\ln$ the natural logarithm. $\mathcal P(X)$ is the set of probability distributions on set $X$. For the lossy Bosonic channel with loss parameter $\eta\in[0,1]$ and power limit $E$, the capacity equals $g(\eta\cdot E)$ where $g(x) =(x+1)\log(x+1)-x\log(x) $ is the Gordon function. We denote by $X\sim \mathcal{N}(\mu, \sigma^2)$ that $X$ is normally distributed with expectation value $\mu$ and variance $\sigma^2$ and $Z\sim\mathcal{N}^{\mathbb C}(\mu, \sigma^2)$ means that $\Re{Z}\sim \mathcal{N}(\Re{\mu}, \sigma^2/2)$ and $\Im{Z}\sim\mathcal{N}(\Im{\mu},\sigma^2/2)$ with $\Re{Z}$ and $\Im{Z}$ being independent. We denote by $\mathds 1_n$ the identity operator of dimension $n$. $A^\dagger$ is the hermitian conjugate of a vector or operator $A$. $\langle x^n,y^n\rangle$ is the inner product of $x^n,y^n$, given by $\sum_{i=1}^nx_i^*y_i$ where $x^*$ is the complex conjugate of $x$. We denote by $\mathbbm i$ the imaginary unit. If no integration bounds are specified, the integral is assumed to be over the entire domain of the function.
    \subsubsection{Definitions}
    The problem is defined as follows:
    We consider a situation where a transmitter sends classical information over a channel with feedback. The input system is modeled by a set $\mathcal X$, the receiver system by a Hilbert space $\mathcal H$, and the feedback system by a Hilbert space $\mathcal K$. 
An \channel is a set of pairs $(\fN_s,\bN_s)$ where each pair consists of two classical-quantum (c-q) channels: One is the memoryless forward channel $\fN_s\in C(\mathcal X,\mathcal H)$, the other a possibly non-memoryless feedback system described by a sequence $\bN_s^n\in C(\mathcal X^n,\mathcal K^{\otimes n})$, where $n\in\mathbb N$ is the block-length. The family of possible channels is labeled by an index $s\in\mathcal{S}$. An illegitimate party may influence the system by making a choice from the set $\mathcal S'=\mathcal S\setminus$\{0\}, otherwise $s = 0$.
\begin{definition}[$(\epsilon,n,\lambda,R)$ code]
An $(\epsilon,n,\lambda,R)$ code $\mathcal C$ under power constraint $E>0$ consists of code-words $x_1^n,\ldots,x_M^n\in\mathcal X^n$ where $M=\lceil2^{n\cdot R}\rceil$ and each codeword $x_m^n$ has entries $\{x_{m,1},\ldots, x_{m,n}\}$, a decoding POVM $\{\Lambda_1,\ldots,\Lambda_M\}$ on $\mathcal H^{\otimes n}$ and detection POVMs $\{\Pi^0_m,\Pi^1_m\}_{m=1}^M$ on $\mathcal K^{\otimes n}$ with average error probability satisfying 
\begin{align}
    \err(\mathcal C):=1-\frac{1}{M}\sum_{m=1}^M\tr(\Lambda_m\fN_0^{\otimes n}(x^n_m))\leq\lambda
\end{align}
and at the same time
\begin{align}
   \min_{s\in\mathcal S'} \frac{1}{2M}\sum_{m=1}^M\left[\tr(\Pi^0_m\bN_0^{n}(x^n_m)) + \tr(\Pi^1_m\bN_s^{n}(x^n_m))\right]\geq1-\epsilon
\end{align}
and let $e(x)$ be a non-negative, non-constant function over the entries of a codeword, then we impose the average energy restriction as 
\begin{align}
    \sum_{j=1}^n  e(x_{m,j})  \leq n E \quad \forall \; m.
\end{align}
\end{definition}
An example of such a channel is the lossy Bosonic channel, where $\mathcal X=\mathbb C$, $\fN_{s}(x)=\ketbra{\sqrt{\eta_s}x}$ and $e(x) = |x|^2$. The number 
$0\leq\eta\leq1$ models the loss of the channel. 
The state $s$ then defines the overall loss parameter experienced by the receiver. We will also give an exemplary model of the channel $\bN$ after defining the capacity/detection region of the model:
\begin{definition}[Rates and Capacity]
A non-negative number $R$ is an achievable \gls{dtr} under power constraint $E>0$ for the \channel with \gls{idr} $D$ if there is a sequence $(\mathcal C_n)_{n\in\mathbb N}$ of $(\epsilon_n,n,\lambda_n,R_n)$ codes such that $\lim_{n\to\infty}\lambda_n=0$ and $\liminf R_n\geq R$ and $\liminf_{n\to\infty}-\tfrac{1}{n}\ln \epsilon_n=D$.  

The quantum rate region of the \channel is the closure of all achievable rate pairs $(D^Q,R^Q)$.

The classical rate region of the \channel is the closure of all achievable rate pairs $(D^C,R^C)$ when the decoding and detection POVMs are homodyne or heterodyne measurements.
\end{definition}
We provide nontrivial points of the rate region for a specific model that aims to describe transmission over an optical fiber.

\begin{SCfigure}
        \includegraphics[width=.26\textwidth]{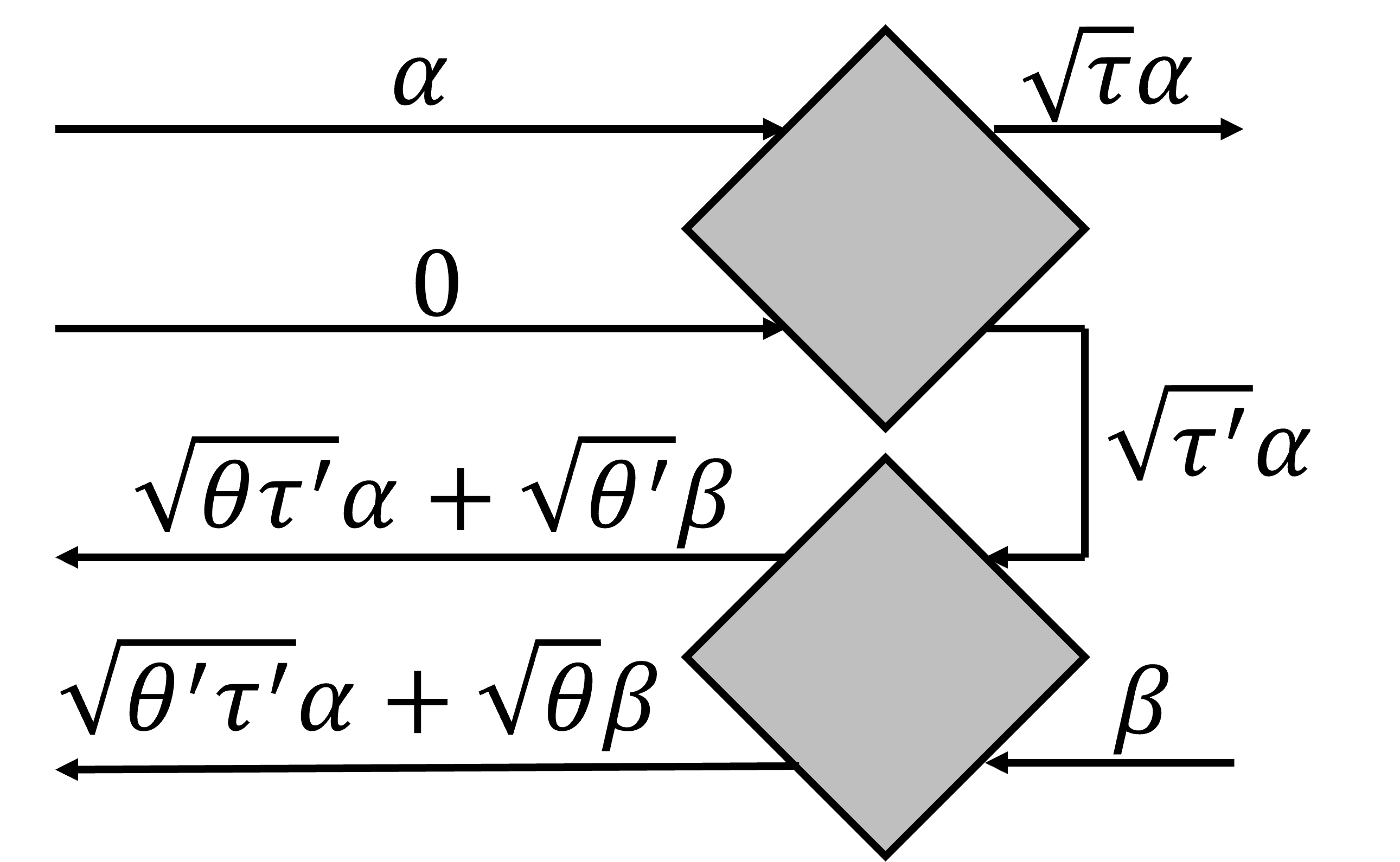} 
        \caption{We model forward propagation and back-scattering with two beam-splitters.  $\tau$ models the propagation loss, while $\theta$ regulates the backward flow (lowest left-pointing port) versus that towards the environment.}
        \label{fig:basic-buliding-block} 

    \end{SCfigure}

\begin{figure}
        \includegraphics[width=.48\textwidth]{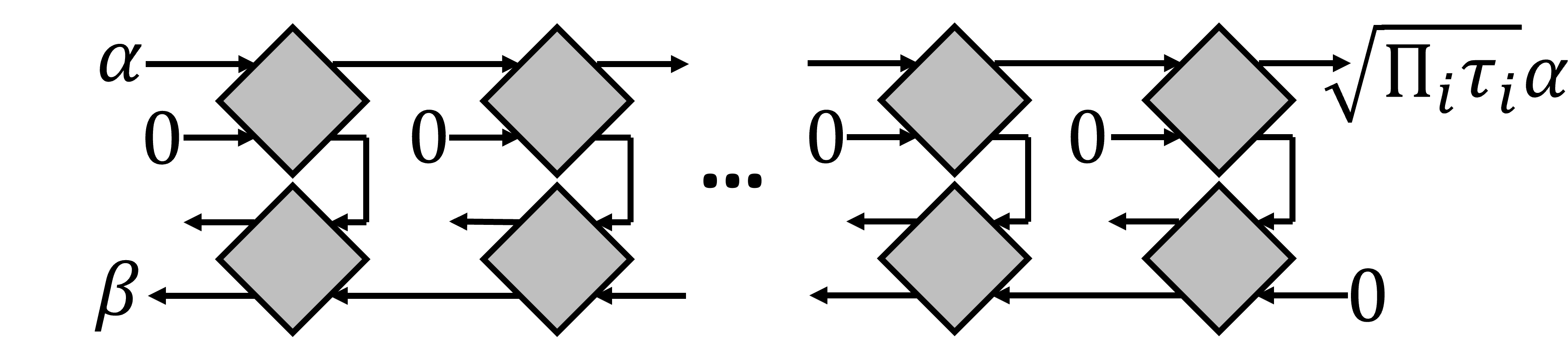}
        \caption{Concatenation of a series of basic building blocks forming a fiber. The value $\sqrt{\prod_i\tau_i}$ can be matched to the loss of fiber. Observe that while each module receives exactly $N=0$ noise photons, the model can in principle incorporate such noise.}
        \label{fig:channel-model}
    \end{figure}

    We continue the specification of our system model by noting that loss affecting a (quantum) signal can be described by the two-mode beam-splitter transformation with transmissivity $\tau$, which acts on coherent states as
    \begin{align}
        BS(|\alpha\rangle|\beta\rangle) = |\sqrt{\tau}\alpha+\sqrt{\tau'}\beta\rangle|\sqrt{\tau}\beta-\sqrt{\tau'}\alpha\rangle.
    \end{align}
    Starting from this basic building block for an optical fiber segment, we model the back-scattering of light in the fiber by the concatenation of $L$ blocks $B_{\tau,\theta}$, where each block $B_{\tau,\theta}$ consists of two beam-splitters. The overall action of a block is that of a $3\times3$ port interferometer, which acts on coherent states as 
    \begin{align}
        B_{\tau,\theta}(|\alpha\rangle|\beta\rangle|0\rangle) = |\sqrt{\tau}\alpha\rangle|a\alpha+b\beta\rangle|\tilde a\alpha+\tilde b\beta\rangle
    \end{align}
    where $a=\sqrt{\tau'\theta'}$, $b=\sqrt{\theta}$, $\tilde a=\sqrt{\tau'\theta}$ and $\tilde b=\sqrt{\tau'\theta'}$. The wiring of the beam-splitters is depicted in Figure \ref{fig:basic-buliding-block}.
    After defining this basic module, all that is left is to concatenate a large number $L$ of the modules as depicted in Figure\ref{fig:channel-model}. Two beam-splitters with transmissivity $\tau$ and $\theta$ are concatenated as in Figure\ref{fig:channel-model}.
    Upon sending a sequence $\alpha^n=\alpha_1,\ldots,\alpha_n$ of coherent states into such a system consisting of $L$ concatenated units, we arrive using the convention $x':=1-x$ at the expression 
    \begin{align} \label{eq:return_state}
        |\beta_t\rangle=\big|\textstyle\sum_{i=1}^L\sqrt{\theta_i'\tau_i'}\big(\prod_{j=1}^i\sqrt{\tau_j\theta_j}\big)\alpha_{t-2i}\big\rangle
    \end{align}
    for the received back-scattered state at time $t$. 
    We can hide the complexity of the expression by writing 
    \begin{align} \label{eq:simple_return_state}
|\beta^n\rangle=\big|A_{n,s}\alpha^n\big\rangle.
    \end{align}
    and $A_{n,s}$ is given by 
    \begin{align*} \label{eq:matrix_A}
    A_{n,s} = \begin{pmatrix}
        a_{1,s}& & & & & &\\
        a_{2,s} & a_{1,s} & &   & & &  \\
        \vdots &&\ddots&&&&\\
        a_{2L,s} & \ldots &\ldots& a_{1,s}  &  &  &  \\
        & \ddots &&&\ddots&& \\
        && a_{2L,s} & \ldots &\ldots& a_{1,s}&
       \end{pmatrix}
\end{align*}
so $\bN^n(\alpha^n) = \ket{A_{n,s}\alpha^n}$. Then, the values $a_{i,s}$ can be related to the specific values of each beam-splitter. In particular, $a_{i,s}=0$ for odd $i$ and 
\begin{equation}
a_{i,s}=\sqrt{\theta'_{i/2}\tau'_{i/2}\Pi_{j=1}^{i/2-1} \theta_j\tau_j} 
\end{equation}
for even $i$.
    In the following, we implicitly assume that $\fN_s$ and $\bN^n_s$ are short-hand for the \gls{lcst} model as introduced above, where the states $s$ are taken from the set $\{(\tau^n,\theta^n)\}$, where $s=0$ models the situation where $\tau_1=\ldots=\tau_n=:\tau$ and $\theta_1=\ldots=\theta_n=:\theta$ for some initial fixed values. The set $\mathcal S'$ is specified as follows: Denote with $T_s(a,b):[0,1]^n\times[0,1]^n$ the operator that maps $(x^n,y^n)$ to $(u^n,v^n)$ where $u_i=x_i$ and $v_i=y_i$ whenever $i\neq s$, in which case $u_i=a$ and $v_i=b$. The set $\mathcal S'$ is then given as $\mathcal S':=\{T_s(a,b)(\tau^n,\theta^n):a\leq\tau,b\in[0,1]\}$. 
    We define $\Pi_i \tau_i:= \eta$ as the total loss parameter of the forward channel $\fN$. 

    This implies that an eavesdropper can tap the system at any point on the line, but will never increase transmissivity for the communication signal.

\section{Results}
\begin{SCfigure}
    
    \includegraphics[width=.3\textwidth]{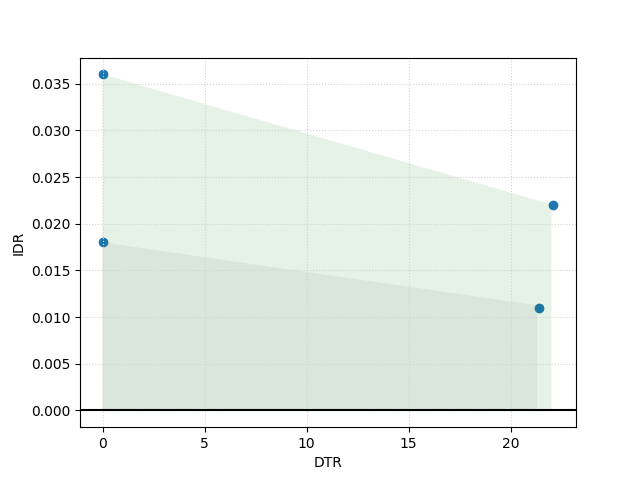}
    \caption{Displayed is an exemplary situation with $L=100$, $\tau_1=\ldots,\tau_{100}=0.99$, $\theta_1=\ldots=\theta_L=0.5$ and $\mathcal S=\{0,1\}$. The value $s=1$ describes a situation where $\tau_{50}=0.4$. $E$ is set to $10^7$.}
    \label{fig:region}
\end{SCfigure}

Our results describe an achievable region of the \gls{lcst} channel. We also describe additional points inside this region that are achievable by utilizing only ``classical'' detection POVMs in the sense that homodyne or heterodyne measurement is applied to each incoming pulse, followed by classical post-processing of the data. In Figure \ref{fig:region} we give a plot for some exemplary parameters of the rate region. The lighter green is the general achievable zone with coherent states and any quantum strategy in the sender and receiver (any POVM) and in darker green the region when we restrict to homodyne/heterodyne measurements at both sender and receiver. In the first case the capacity is given by the Holevo capacity and in the second one by the Shannon capacity. Throughout, we assume one fixed but arbitrary value $E>0$ is set for the average power constraint of the transmitter.
\begin{theorem}\label{thm:achievable-rate-region}
    Let $\mathcal R$ be the set of pairs $(R,D)$ such that $R$ is an achievable \gls{dtr} and $D$ is an achievable \gls{idr}, given by $\lim_{n\xrightarrow[]{}\infty}\-\tfrac{1}{n}\ln \epsilon$, where $\epsilon$ is an error probability. To every $\mu\in\mathcal P(\mathbb C)$ satisfying $E(\mu):=\int e(\alpha) d\mu(\alpha)\leq E$ and to every $s\in\mathcal S'$ we define $D_s(\mu)=\lim_{n\to\infty}\tfrac{1}{n}\ln\mathbb E_\mu \epsilon$. Then
    \begin{align}
        \{(\chi(p;\mathcal \fN_0),\min_{s\in\mathcal S'}D_s(\mu)):E(\mu)\leq E\}\subset\mathcal R.
    \end{align}
\end{theorem}
\begin{lemma}[Maximum Intrusion 
Detection Rate] \label{lem:maximum-idr}
    Let the maximum \gls{idr} be defined as $D^Q_{\mathrm{max} D}:=\max\{D:(R,D)\in\mathcal R\}$.  It holds $D^Q_{\mathrm{max} D} = E \min_{s\in\mathcal{S}'}\left(\sum_{k=1}^{2L} c_{k,s}\right)^2$. Thus, the point $(0,D^Q_{\mathrm{max} D})$ is achievable. 
\end{lemma}
\begin{lemma}[Optimal Classical Intrusion Detection Rate]
\label{lem:maximum-idr-classical}
    The statement of Lemma \ref{lem:maximum-idr} holds for the classical detection strategy with $D^Q_{\max D}$ replaced by $D^C_{\max D}:=\frac{1}{2}D^Q_{\max D}$.
\end{lemma}

Lastly, we characterize an \gls{idr} corresponding to the point where the \gls{dtr} is maximized:
\begin{lemma}[Achievable Intrusion Detection Rate under Maximum Rate]
\label{lem:random-coding-idr}
The number $R^Q_{\mathrm{max} R} = g(\eta E)$ is the supremum over all achievable \gls{dtr} for the \gls{lcst} channel. The \gls{idr} given by $D_{\max R}^Q = \min_s \int_0^1 \log\left(E\sum_{m,l=1}^{2L}c_{m,s}c_{l,s}e^{2\pi i(m-l)x}+1\right)dx $ is achievable under $R^Q_{\mathrm{max} R}$. Thus, $(R^Q_{\mathrm{max} R}, D_{\max R}^Q)\in \mathcal{R}$.
\end{lemma}

\section{Proofs}
 We will apply the Helstrom bound \cite{Helstrom_original} to obtain an upper bound on the \gls{idr}. The Helstrom bound for pure states with equal prior probabilities gives the minimum average error probability to distinguish between pure states $\ket{\alpha}$ and $\ket{\beta}$ as 
 \begin{equation}
 P_e^*= \frac{1}{2}\left(1-\sqrt{1-|\braket{\alpha}{\beta}|^2}\right).
 \end{equation}
 Note that the Helstrom bound is achievable for binary hypothesis testing. Still, since in this problem one of the hypotheses is composite, even the worst Helstrom bound between the possible hypotheses might not be achievable. However, we give a strategy that achieves the asymptotic scaling of Helstrom bound and thus the optimal \gls{idr} for all possible $s\in \mathcal{S'}$. 
  To detect an eavesdropper, we need to solve a hypothesis testing problem where the null hypothesis is given by  $\ket{A_{n,0}\alpha^n}$ and the alternative hypothesis is composite, $\ket{A_{n,s}\alpha^n}$ for some $s\in\mathcal{S}'$. We generally define the POVM to distinguish between $\ket{\alpha^n}$ and $\ket{\beta_s^n}$ as 
  \begin{equation}
  \label{eq : povm}
  \begin{aligned}
      \Pi_0 &=D(-\alpha^n)^\dagger\ketbra{0^n}D(-\alpha^n) 
      \\\Pi_1 &= \mathds 1_n- D(-\alpha^n)^\dagger\ketbra{0^n}D(-\alpha^n)
  \end{aligned} 
  \end{equation}
  which achieves the optimal \gls{idr}. This POVM is constructed by displacing the known state to 0 and then doing a 0 or not photon number measurement. The error probability is
\begin{align}
\label{eq:error-prob}
P_e =& \frac{1}{2}\big[\tr(\ketbra{\beta^n_s-\alpha^n}\ketbra{0^n}) +\\ \nonumber&\tr( (\mathds 1_n- \ketbra{0^n})\ketbra{0^n})\big]\\
    =&\frac{1}{2}|\bra{\beta^n_s-\alpha^n}\ket{0^n}|^2    
\end{align}
where $|\bra{\alpha^n}\ket{\beta^n}|^2= \Pi_{i=1}^n \exp{-\abs{\alpha_i - \beta_i}^2}$ is the inner product of two $n$-mode coherent states. Given input $\alpha^n$, $\bN_s(\alpha^n) = A_{n,s}\alpha^n$ and $\bN_0(\alpha^n) = A_{n,0}\alpha^n$ and the probability of error for this POVM is $P_e = \frac{1}{2}|\bra{A_{n,s'}\alpha^n}\ket{A_{n,0}\alpha^n}|^2$. 
 where we define $C_{n,s} = A_{n,s} - A_{n,0}$.
 \begin{equation}
 \label{eq:error-prob-exp}
     P_e = \frac{1}{2}\exp{-\langle C_{n,s}\alpha^n, C_{n,s} \alpha^n\rangle} = \frac{1}{2}\exp{-\langle \alpha^n, G_{n,s} \alpha^n\rangle} 
 \end{equation}
 As $a_{i,s}\geq a_{i,0}\forall i$, $c_{i,s}\geq0$ and $C_{n,s}$ remains as a lower triangular matrix. 
 Finally, we define the matrix $G_{n,s} = C^\dagger_{n,s} C_{n,s}$, which is a real Hermitian Toeplitz matrix. That is, diagonals have constant value and $G_{n,s}^\dagger = G_{n,s}$. In particular, diagonal $j$ with $j=0$ being the main diagonal, has value $g_{i,s}=\sum_{k=0}^{2L}c_{k,s}c_{k+i,s}$ and $c_{k,s}=0,g_{k,s}=0\forall k>2L$. Then the Toeplitz matrix $G_{n,s}$ is generated by
\begin{equation}
\label{eq:generating-function}
    f(\xi, s) =  \sum_{k=-\infty}^\infty g_{k,s} e^{\mathbbm{i}k\xi} = g_{0,s}+ \sum_{k=1}^{2L-1} g_{k,s}(e^{\mathbbm{i}k\xi}+e^{-\mathbbm{i}k\xi}) 
\end{equation}
in the sense that $g_{k,s} =\frac{1}{2\pi}\int_0^{2\pi} f(\xi, s)e^{-\mathbbm i\xi k}d\xi$.
This formulation allows us in the following to use theorems proved for semi-infinite Toeplitz matrices\cite{toeplitz_matrices}\cite{toeplitz-circulant-review}. 

    \subsection{Concentration of measure for IDR of Random Codes}
\begin{IEEEproof}[Proof of Theorem \ref{thm:achievable-rate-region}]
    Let $\mu\in\mathcal P(\mathbb C)$ satisfy $\int e(\alpha)d\mu(\alpha)\leq E$. By the random code construction \cite{holevoBook} this distribution it is known that a method by which we sample $M$ individual code-words $\alpha^n$ i.i.d. according to the product measure $\mu^{\otimes n}$ then the probability that the respective code-book $\mathcal C$ satisfies $\tfrac{1}{n}\log M\geq \chi(p;\fN)-\delta$ while achieving $\err(\mathcal C)\leq2^{-\delta\cdot n}$ is lower bounded by $1-2^{-\delta\cdot n}$. Here, $\delta>0$ is arbitrary, and the statement holds for all large enough $n\in\mathbb N$. 

    Consider a signal-dependent POVM $\{\Pi^0_{x^n},\Pi^1_{x^n}\}$. Define for every $s\in\mathcal S'$ a random variable
    \begin{align}
       P_{e}^s(x^n):=\frac{1}{2}\left[\tr(\Pi^1_{x^n}\bN_0^{n}(x^n)) + \tr(\Pi^0_{x^n}\bN_s^{n}(x^n))\right], 
    \end{align}
    and let for every $s\in\mathcal S'$
    \begin{align}
        \mathbb EP_e^s:=\sum_{\mathcal C}p(\mathcal C)\frac{1}{M}\sum_{m=1}^MP^s_{\epsilon}(X^n_m).
    \end{align}
    According to Markov's inequality it holds $\mathbb P(D_s > t)\leq t^{-1}\mathbb ED_s$. By setting $t=2^{n\cdot\delta}\cdot\mathbb EP_e^s$ and using a union bound we therefore arrive at
    \begin{align}
        \mathbb P(\min_{s\in\mathcal S'} P_e^s - 2^{n\cdot\delta}\cdot\mathbb EP_e^s>0)\leq |\mathcal S'|2^{-n\cdot\delta}.
    \end{align}
    This implies that for every $\delta>0$ the existence of sequences of codes achieving asymptotically both detection rate $\min_{s\in\mathcal S'}\lim_{n\to\infty}-\tfrac{1}{n}\log \mathbb EP_e^s + \delta$ and communication rate $\chi(p;\fN)-\delta$. By letting $\delta$ go to zero, the theorem is proven.
\end{IEEEproof}
     
     \subsection{Optimal detection strategy and detection exponent}\label{subsec:optimal-detection-and-exponent}
    
\label{proof : optimal-det}
\begin{IEEEproof}[Proof of Lemma \ref{lem:maximum-idr}]
 Optimal \gls{idr} is achieved by the sequence given by
\begin{equation}
 \arg \min_{\alpha^n}\max_{s\in\mathcal S'} -\langle \alpha^{n} , G_{n,s}\alpha^n\rangle
\end{equation}
where $\langle\alpha^n,\alpha^n\rangle \leq nE$ due to the energy constraint, as this condition is equivalent to minimizing the inner product between possible output states and the error probability $P_e^*$ is monotone in the inner product.
All achievable \gls{idr} are upper bounded by 
\begin{equation}
\label{eq : limit}
    D^Q_* = \lim_{n\xrightarrow[]{}\infty}-\frac{1}{n}\ln{P_e^*} 
\end{equation}
where $P_e^* = \frac{1}{2}\left(1-\sqrt{1-|\braket{A_{n,0}\alpha^n}{A_{n,s}\alpha^n}|^2}\right)$ and $|\braket{A_{n,0}\alpha^n}{A_{n,s}}|^2 = e^{-\langle \alpha^n,G_{n,s}\alpha^n\rangle}$.
By \cite[Corollary 4.2]{toeplitz-circulant-review}
\begin{equation}
\label{eq : limit-eigenvalue}
    \lim_{n\xrightarrow[]{}\infty}\max_{\alpha^n} \frac{\langle \alpha^n, G_{n,s} \alpha^n\rangle}{n} = \lim_{n\xrightarrow[]{}\infty} \frac{1}{n}\lambda^{\max}_{n,s} = E\sup_\xi f(\xi, s)
\end{equation}
where $\lambda^{\max}_{n,s}$ is the largest eigenvalue of $G_{n,s}$ for size $n$ and eavesdropping parameter $s$. 
Then, there exists an input such that for large enough $n$ the optimal error probability behaves as 
\begin{align}
    P_e^*= \frac{1}{2}\left(1-\sqrt{1-e^{-n E\min_sf(0,s)}}\right)
\end{align}
and the limit in Equation $\ref{eq : limit}$ can be shown to converge to $D_*^Q = \min_sf(0,s)$. Now we evaluate the performance of the POVM given in Equation $\ref{eq : povm}$, that its \gls{idr} is given by
\begin{equation}
    D_{\max D}^Q = \lim_{n\xrightarrow[]{}\infty}-\frac{1}{n}\ln P_e = \min_{s\in\mathcal{S}'}\lim_{n\xrightarrow[]{}\infty}\frac{\langle \alpha^{n} , G_{n,s}\alpha^n\rangle}{n} 
\end{equation}
where $P_e$ is given in Equation \ref{eq:error-prob-exp}. By Equation  \ref{eq : limit-eigenvalue}, this is again given by  
 \begin{equation}
     D_{max D}^Q = E\min_sf(0,s) = D_*^Q
 \end{equation}
and thus, this strategy achieves the optimal \gls{idr}.
Since $c_{k,s}\geq 0$ and by the form of $f(\xi,s)$ (Equation \eqref{eq:generating-function}), it is straightforward to see that the maximum is achieved by $\xi= 0$ independently of $s$. $\min_sf(0,s)$ and we can simplify the expression for $D_{\max D}^Q$
\begin{align}
    D_{\max D}^Q &=\min_{s\in\mathcal{S}'} E \left(g_{0,s}+ 2\sum_{k=1}^{2L-1}g_{k,s}\right)^2\\ &=\min_{s\in\mathcal{S}'} E\left(\sum_{k=1}^{2L} c_{k,s}\right)^2
\end{align}
where a factor $E$ is added as the vector is normalized to $E$ due to the energy constraint. 
\end{IEEEproof}
\begin{IEEEproof}[Proof of Lemma \ref{lem:maximum-idr-classical}] \label{sec:proof_DmaxD}
The strategy described above minimizes the overlap between the expected return state when $s=0$ and the actual state in case that $s\neq0$. Discarding the first $2L$ transmissions gives us $n-2L$ copies of the coherent state 
\begin{align}
    |\alpha_s\rangle = |\sqrt{E}\sum_{i=1}^{2L}c_{i,s}\rangle 
\end{align}
xSince we know the phase of the state as all $c_{i,s}>0$, we can use homodyne detection. When measuring the $x$-quadrature operator on the coherent state $\ket{\alpha_s}$ with $\alpha_s \in \mathds R$, the probability of getting the outcome $\phi$ is distributed according to $\mathcal{N}(\sqrt{2}\alpha_{s},\frac{1}{2})$\cite{brask2022gaussian}. Therefore, the task is a classical hypothesis testing problem of two normal distributions with known, identical variance. In this case, the asymptotic exponent of the average error is given by the Chernoff information \cite[Chapter 11.9]{cover2012elements} of the distributions corresponding to the channel states $0$ and $s$. The Chernoff information of two distributions $f_0(x)$ and $f_1(x)$ is 
\begin{align} \label{eq:chernoff_info}
    C(f_0,f_1) = \max_{\lambda \in [0,1]} - \ln\int f_0(x)^\lambda f_1(x)^{1-\lambda} dx.
\end{align}
For our two real Gaussian distributions $\mathcal N (0,\frac{1}{2})$ and $\mathcal N (\sqrt{2}\alpha_s,\frac{1}{2})$ we get
\begin{align}\label{eq:chernoff_info_gaussians}
    C\left( \frac{1}{\sqrt{\pi}} e^{-x^2}, \frac{1}{\sqrt{\pi}} e^{-(x-\sqrt{2}\alpha_s)^2} \right) = \frac{\alpha^2_s}{2}.
\end{align}
Therefore the asymptotic error exponent for distinguishing channel states $0$ and $s$ in this way is $\frac{E}{2} \left( \sum_{i=1}^{2L} c_{i,s} \right)^2$. Note that this only applies to testing two simple hypotheses, so when we don't know which channel state to test for, this result does not apply directly. However, we can always apply the test corresponding to the smallest IDR and achieve a detection rate $\geq \min_{s\in\mathcal{S}'} \frac{E}{2} \left( \sum_{i=1}^{2L} c_{i,s} \right)^2$. This is because by doing so, the probability of committing an error of the first kind is the same for all $s \in \mathcal S$, while the probability for an error of the second kind is upper bounded by the one corresponding to the smallest IDR in all cases. Therefore 
\begin{equation}
    D_\mathrm{\max D}^C = \min_{s\in\mathcal{S}'} \frac{E}{2} \left( \sum_{i=1}^{2L} c_{i,s} \right)^2 = \frac{1}{2}D^Q_\mathrm{\max D}
\end{equation}.

\end{IEEEproof}
        
\subsection{Maximum IDR under optimal communication}            \begin{IEEEproof}[Proof of Lemma \ref{lem:random-coding-idr}]
It is known \cite{gaussian_distribution_is_optimal} that for a lossy bosonic channel $\alpha \rightarrow \ket{\sqrt{\eta} \alpha}$ with average power constraint $\sum_{i=1}^n \abs{\alpha}^2 \leq nE$ the optimal \gls{dtr} $R = g(\eta E)$ is achieved by a random code over a complex normal distribution $\mathcal{N}^\mathbb C (0,E)$.
 To find the eavesdropper detection rate, we consider the expectation value over the random codebook definition and the selection of the message $m$ as stated in Theorem \ref{thm:achievable-rate-region}.  Applying the POVM given in \eqref{eq : povm} (with the displacement being set according to the sent state) we give a lower bound for the \gls{idr}. The error probability is given by
\begin{equation}
 P^\mathcal{N}_e =\exp{-\langle \alpha^n,G_{n,s}\alpha^n\rangle}  
\end{equation}
where the superindex $\mathcal{N}$ is used to remark that $\alpha^n$ is a random variable where each component is distributed as $\mathcal{N}^\mathbb C(0,E)$ and uncorrelated to the others, thus the probability of error is also a random variable. 

To calculate the expectation value and show that for large $n$ it converges we use known results of $n$-dimensional Gaussian integrals. Generally, it can be shown that 
\begin{equation}
\label{gaussianintegral}
    \int d((\phi^l)^\dagger, \phi^l) \exp{-(\phi^l)^\dagger M_l\phi^l} = \det M_l^{-1}
\end{equation}
where $M_l$ is an $l \cross l$ hermitian matrix such that $\Re{M_l}>0$, $\phi^l$ and $(\phi^l)^\dagger$ are vectors of length $l$ with components given by the integration variables $\phi_j^*$ and $\phi_j$ \cite[Equation 3.18]{gaussianintegrals}. We use the short-hand notation $d((\phi^l)^\dagger, \phi^l)=\Pi_{i=1}^N\frac{1}{\pi}d\phi_id\phi_i^*$. We integrate over the variables and their conjugate, which is equivalent to integrating over the real and imaginary part, up to a constant. 

Then, we can write our expectation value of the error probability $P_e^\mathcal{N}$ as 
\begin{equation}\label{eqn:expected-error-probability}
    \mathbb E P_e^\mathcal{N} =   \int p(\alpha^n)\exp{-(\alpha^n)^\dagger G_{n,s}\alpha}d\alpha^n d(\alpha^n)^*.
\end{equation}
Since $p(\alpha^n)=\prod_{i=1}^n\frac{1}{E\pi}e^{-|\alpha_i|/E} = \frac{1}{(E\pi)^n}e^{-(\alpha^n)^\dagger \mathds 1_n \alpha^n/E}$ \eqref{eqn:expected-error-probability} can be rewritten as an integral of the form
\begin{equation}
    \mathds{E} P_e^\mathcal{N}= \frac{1}{E^n}\int d((\phi^n)^\dagger,\phi^l) \exp{-(\phi^n)^\dagger\left(G_{n,s}+\frac{\mathds{1}_n}{E}\right)\phi^n}.
\end{equation}
Since $G_{n,s}$ is positive and hermitian by construction as $C_{n,s}^\dagger C_{n,s}$, we can apply \eqref{gaussianintegral}. This yields 
\begin{equation}
\label{expectationqexp}
    \mathds E P_e^\mathcal N= \frac{1}{E^n}\det\left(G_{n,s}+\frac{\mathds 1}{E}\right)^{-1} = \Pi_{i=1}^n \left(E\lambda_{i,n,s}+1\right)^{-1}
\end{equation}
where the notation $\lambda_{i,n,s}$ for the eigenvalues of $G_{n,s}$ is used to remark that the eigenvalue $i$ depends on the size of the matrix $n$ and the value $s$.
Upon taking the limit $n\xrightarrow[]{}\infty$, $\mathds E P_e^\mathcal{N}$ diverges to 0 (in the sense that $\sum_{j=0}^\infty\ln(\frac{1}{E\lambda_{j,n,s}+1})=-\infty$, as all terms are negative and upper-bounded), thus the error probability goes to 0. The asymptotic exponent $D_{\max R}^Q$ is then given by 
\begin{equation}
\label{eq:function-eigenvalues}
    \min_{s\in\mathcal{S}'}\lim_{n\xrightarrow[]{}\infty}-\frac{1}{n}\ln\left(\mathds E D\right) = \min_{s\in \mathcal{S'}}\lim_{n\xrightarrow[]{}\infty}\frac{1}{n}\sum_{i=1}^n\ln\left(E\lambda_{i,n,s}+1\right)
\end{equation}

Finally, we apply one of the Szegö limit theorems \cite{asymptotic-spectra}\cite{grenandertoeplitz}. This theorem states that for a function $F$ on the eigenvalues of an infinite Toeplitz matrix generated by $f(\xi)$,
\begin{equation}
    \lim_{n\xrightarrow[]{}\infty}\frac{1}{n}\sum_i F(\lambda_{i,n,s}) = \frac{1}{2\pi}\int_0^{2\pi}F(f(\xi))d\xi
\end{equation}
as long as $F(\cdot)$ is continuous on the interval $(\sup_\xi f(\xi), \inf_\xi f(\xi))$. Applying this result to \eqref{eq:function-eigenvalues}  we obtain 
\begin{equation}
    D_{\max R}^Q = \min_{s\in\mathcal{S}'} \frac{1}{2\pi}\int_0^{2\pi} \ln\left(Ef(\xi,s)+1\right)d\xi
\end{equation}

where $f(\xi, s)$ is given in Equation \eqref{eq:generating-function}, changing back to dependency on $c_i,s$, we obtain the expression
\begin{equation}
\label{error-exp}
    D_{\max R}^Q = \min_{s\in\mathcal{S}'} \int_0^1 \ln\left(E \sum_{m,l=1}^{2L}c_{m,s}c_{l,s}e^{2\pi \mathbbm{i}(m-l)x}+1\right)dx
\end{equation}
which is hard to solve analytically but can easily be evaluated numerically given a specific set of values. 
\end{IEEEproof}

\section{Conclusions}
In this work, a new model for simultaneous communication and eavesdropper detection has been introduced. We showed that a region of \gls{idr} and \gls{dtr} rate can be constructed. We determined the maximum value for the \gls{idr} when the goal is to optimize for detection and an achievable \gls{idr} when optimizing for communication. We also compared optimal detection to the one achievable using only homodyne/heterodyne measurements, finding that its \gls{idr} is by a factor of 2 below the optimal value when these conventional techniques are used. Detection performance is directly proportional to the input energy per pulse. We remark that our work's main contribution is considering Joint Communication and Sensing in a quantum channel with memory and in infinite dimensions. Moreover, we consider composite binary hypothesis testing instead of multi-hypothesis testing.  
Future work must model the optimal regions for identifying or learning the system parameter $s$, which includes the eavesdropper location. Other directions could include using other families of states for detection, such as squeezed states, which are known to perform better in parameter estimation tasks 
, and considering other channel models, such as an analog for free space transmission or including other optical fiber effects.  
\bibliography{IEEEabrv, bib}
\end{document}